\documentclass[12pt]{article}
\usepackage{epsfig,amsfonts,amssymb,amsbsy,amsbsy}
\def\cvp{\raise 2pt\hbox{,}}

\def\tr{\mathop{\rm tr}\nolimits}

\def\plb#1#2#3{{\it Phys.\ Lett.\ }{\bf B #1} (#2) #3}
\def\npb#1#2#3{{\it Nucl.\ Phys.\ }{\bf B #1} (#2) #3}
\def\prl#1#2#3{{\it Phys.\ Rev.\ Lett.\ }{\bf #1} (#2) #3}
\def\jhep#1#2#3{{\it J. High Energy Phys.\ }{\bf #1} (#2) #3}

\def\cmp#1#2#3{{\it Comm.\ Math.\ Phys.\ }{\bf #1} (#2) #3}
\def\pr#1#2#3{{\it Phys.\ Rep.\ }{\bf #1} (#2) #3}
\def\jmp#1#2#3{{\it J.\ Math.\ Phys.\ }{\bf #1} (#2) #3}
\def\ijmpa#1#2#3{{\it Int.\ J.\ Mod.\ Phys.\ }{\bf A #1} (#2) #3}
\def\mpla#1#2#3{{\it Mod.\ Phys.\ Lett.\ }{\bf A #1} (#2) #3}
\def\jpa#1#2#3{{\it J.\ Phys.\ }{\bf A #1} (#2) #3}
\begin{document}
%
%\pagenumbering{roman}
%
\pagestyle{empty}
{\parskip 0in
\hfill PUPT-1998

\hfill NEIP-01-009

\hfill LPTENS-02/11

\hfill hep-th/0202205}

\vfill
\begin{center}
{\LARGE Non-perturbative double scaling limits}
%\medskip

\vspace{0.4in}

Frank F{\scshape errari}{\renewcommand{\thefootnote}{$\!\!\dagger$}
\footnote{On leave of absence from Centre 
National de la Recherche Scientifique, Laboratoire de Physique 
Th\'eorique de l'\'Ecole Normale Sup\'erieure, Paris, France.}}\\
\medskip
{\it Institut de Physique, Universit\'e de Neuch\^atel\\
rue A.-L.~Br\'eguet 1, CH-2000 Neuch\^atel, Switzerland\\
and\\
Joseph Henry Laboratories\\
Princeton University, Princeton, New Jersey 08544, USA}\\
\smallskip
{\tt frank.ferrari@unine.ch}
\end{center}
\vfill\noindent
Recently, the author has proposed a generalization of the matrix and 
vector models approach to the theory of random surfaces and polymers. 
The idea is to replace the simple matrix or vector (path) integrals by gauge 
theory or non-linear $\sigma$ model (path) integrals. We explain how 
this solves one of the most fundamental limitation of the classic 
approach: we automatically obtain non-perturbative definitions in 
non-Borel summable cases. This is exemplified on the simplest 
possible examples involving ${\rm O}(N)$ symmetric non-linear
$\sigma$ models with $N$-dimensional target spaces, for which we 
construct (multi)critical metrics. 
The non-perturbative definitions of the double scaled, manifestly positive, 
partition functions rely on remarkable identities involving (path) integrals.
\vfill
\begin{flushleft}
February 2002
\end{flushleft}
\newpage\pagestyle{plain}
\baselineskip 16pt
\setcounter{footnote}{0}
%La ligne suivante permet la numerotation des equations par section
%\renewcommand{\theequation}{\thesection.\arabic{equation}}
% La ligne suivante fait pareil mais dans le format amsmath
%\numberwithin{equation}{section}
%\tableofcontents\pagenumbering{arabic}
%
\section{Motivations and example}

Finding a general non-perturbative definition of string theory remains 
one of the 
most challenging problem in theoretical physics. One may hope that such a 
definition will automatically follow from an understanding of the basic 
principles of the theory and/or of quantum gravity, but for the moment we 
must rely on the `universal' or `unique' nature of the theory in order to 
gain an insight \cite{polbook}. Historically, the first serious attempt at a 
non-perturbative approach was through the study of matrix models 
\cite{BIPZ} and the double scaling limits \cite{BK}. The idea \cite{DK}
was that near critical points, very large Feynman diagrams dominate, and 
thus in the 't Hooft representation \cite{tHooft} the matrix theory
reduces to a sum over {\it continuous} world-sheets. Comprehensive 
discussions of these subjects can be found in \cite{rev1,rev2,rev3},
and short introductions are in recent papers by the author
\cite{fer1,fer2}. 
Unfortunately, the detailed investigations of the matrix integrals revealed 
that a non-perturbative definition of the most interesting theories, the 
unitary models which have non-Borel summable partition functions, could 
not be achieved. This point is discussed in details for example in Section 
7 of \cite{rev1}. Typically, the matrix model approach yields a 
differential equation, called the string equation,
that determines unambiguously the perturbative, 
asymptotic series expansion, but that has several solutions differing by 
exponentially suppressed terms. For example, the string equation for the 
simplest critical point, that corresponds to pure two-dimensional gravity, 
is the Painlev\'e I differential equation \cite{BK}
\begin{equation}
\label{se1}
z = u^{2}(z) + {1\over 3} u''(z)\, ,
\end{equation}
where the closed string coupling constant is
$\kappa = z^{-5/4}$, and the connected partition function
\begin{equation}
\label{Zexp}
W = \sum_{h=0}^{H} W_{h}\, \kappa^{2h-2} + {\cal O}(\kappa^{2H})
\end{equation}
is such that $W''(z) = u(z)$. Equation (\ref{se1}) implies a recursion 
relation that determines all the coefficients $W_{h}$ once the sphere 
contribution $W_{0}$ is known.
However, at large $h$, the coefficient $W_{h}$ goes
like $(2h)!$, and thus the expansion (\ref{Zexp}) is not Borel summable and 
does not define a unique function. It turns out that solutions to 
(\ref{se1}) with the asymptotic expansion (\ref{Zexp}) are parametrized by 
an arbitrary real number, and differ at small $\kappa$  by terms of order 
$\exp (-4\sqrt{6} /(5\kappa))$ \cite{BK, rev1, rev3}. Those crucial 
non-perturbative contributions remain unknown. It is actually possible to 
understand in an elementary way why the matrix model fails to provide a 
non-perturbative definition. The model is defined by the integral over 
hermitian $N\times N$ matrices $M$,
\begin{equation}
\label{matint}
\int\! d^{N^{2}}\! M\,\exp\Bigl(
-{1\over 2} \tr M^{2} - {g\over {N}} \tr M^{4}\Bigr)\, ,
\end{equation}
and it turns out that
the critical point lies at a negative value of $g$, for which the 
integral (\ref{matint}) is divergent. The same critical point 
could be obtained by starting from more general matrix integrals
\begin{equation}
\label{matint2}
\int\! d^{N^{2}}\! M\,\exp\Bigl(-{N\over g}\tr U(M)\Bigr)\, ,
\end{equation}
with an arbitrary potential $U(M)$, but the fact is that
the pure gravity, or any other unitary critical point, is always unstable. 
Shortly after the discoveries of \cite{BK}, it was realized \cite{polym} 
that the same procedure could be applied to theories of polymers, by 
replacing matrix integrals like (\ref{matint2}) by ${\rm O}(N)$ symmetric
vector integrals like
\begin{equation}
\label{vecint}
\int\! d^{N}\vec V\, \exp\Bigl( -{N\over g}\, U({\vec V}^{2})\Bigr)\, .
\end{equation}
In addition to their intrinsic interest, the polymer integrals are useful 
toy models for the more complicated string theories. An interesting 
aspect is that higher dimensional vector path integrals can be easily studied 
\cite{polymhD}, in contrast to the matrix case. Works on the vector models 
are reviewed in \cite{polymrev}. However, the basic problem 
stressed above, that non-perturbative results cannot be obtained in the 
most interesting non-Borel summable cases, plagues the vector integrals in 
the same way as it plagues the matrix integrals, and for the same reasons. 
In spite of several attempts over the years (see in particular 
\cite{fermat} and references therein), this fundamental drawback of the 
matrix (or vector) models approach could never be satisfactorily solved.
However, very recenty, the author made a simple proposal that 
automatically overcome the difficulty \cite{fer3,fer4,fer1,fer2}.
The purpose of the present note is to show explicitly how this proposal
works in the simplest possible examples.

The idea
is to replace the simple matrix (\ref{matint2}) or vector (\ref{vecint}) 
(path) integrals by some gauge theory or non-linear $\sigma$ models 
(path) integrals respectively. The non-trivial result is 
that analogues of the Kazakov critical points \cite{DK} exist in those 
cases as well. For two dimensional non-linear $\sigma$ models, it was 
shown in \cite{fer3} that mass terms for the would-be Goldstone 
bosons could be adjusted to critical values, and double scaling limits 
defined. It was also argued at length in \cite{fer3} that mass (or more 
generally potential) terms in 
non-linear $\sigma$ models are very similar to Higgs vevs in gauge 
theories. And indeed in \cite{fer4,fer1} is was shown that in 
four dimensional supersymmetric gauge theories, the adjoint
Higgs vevs moduli can also be adjusted 
to critical values and double scaling limits defined, with the 
Argyres-Douglas singularities \cite{AD} playing the r\^ole of the Kazakov 
critical points. This yields four dimensional non-critical (or five 
dimensional critical) string theories \cite{fer1}.
In all those examples, {\it the double scaling limits are 
always non-perturbative, because the original integrals are convergent for 
all values of the parameters.}

To illustrate this point, we will focus in the following
on $D=0$, $1$ or $2$ dimensional
non-linear $\sigma$ model examples, akin to the model 
studied in \cite{fer3}. In those cases, the large $N$ 
expansion is a loop expansion in the dual 
representation of Feynman diagrams and, as reviewed in \cite{fer2}, this 
implies that the double scaled theories are field theories themselves 
(in contrast with the gauge theory case which yields string theories).
In spite of this considerable simplification, the basic difficulty 
remains the same. For example, the simplest
non-Borel summable double scaled field theoretic partition function we will 
encounter is
\begin{equation}
\label{int1}
e^{W(\kappa)} = Z(\kappa) =
\int\! {dx\over\sqrt{2\pi}}\,e^{-x^{2}/2 +\kappa x^{4}/4} .
\end{equation}
Obviously, this integral diverges, but admits a well-defined perturbative 
expansion
\begin{equation}
\label{Zexp1}
Z(\kappa) = 1 + \sum_{k=1}^{K} {(4k-1)!\over 2^{4k-1} k! 
(2k-1)!}\,\kappa^{k} + {\cal O}(\kappa^{K+1}).
\end{equation}
The integral representation (\ref{int1}) makes `unitarity' obvious: each 
diagram will have a positive weight. This has the usual consequence that
all the coefficients in the series expansion for $W = \ln Z$ are 
positive, and the series is not Borel summable. 
The `string equation,' analogue to 
(\ref{se1}), is the Schwinger-Dyson equation 
for (\ref{int1}). It takes a simple form when written in terms of $Z$,
\begin{equation}
\label{pe1}
16\kappa^{2}\, Z'' + 4(8\kappa -1)\, Z' + 3\, Z = 0\, .
\end{equation}
With the condition $Z=1+{\cal O}(\kappa)$, this equation implies 
the asymptotic expansion (\ref{Zexp1}), in the same way as (\ref{se1}) 
implies a unique expansion (\ref{Zexp}). And also in strict parallel 
with (\ref{se1}), there is a one parameter family of solution to 
(\ref{pe1}) with the correct asymptotic behaviour. It can be written 
in terms of modified Bessel functions,
\begin{equation}
\label{zt1}
Z_{\theta}(\kappa) = {\sqrt{\pi}e^{-1/(8\kappa)}\over 4\sqrt{\kappa}}  \left[
I_{1/4}\Bigl({1\over 8\kappa}\Bigr) + I_{-1/4}\Bigl({1\over 8\kappa}\Bigr)
+ \theta \Bigl[I_{1/4}\Bigl({1\over 8\kappa}\Bigr) - 
I_{-1/4}\Bigl({1\over 8\kappa}\Bigr) \Bigr]\right] .
\end{equation}
The combination $I_{1/4}-I_{-1/4}$ is a purely non-perturbative 
contribution to $Z$. In agreement with the high order behaviour of 
(\ref{Zexp1}), it yields terms proportional to $\exp(-1/(4\kappa))$ 
that remain unknown.

In the standard matrix case \cite{rev1}, equations (\ref{se1}) and 
(\ref{Zexp}), we have no hint to the 
value of the $\theta$ parameter. It is actually far from 
being obvious that there exists a `correct' solution to (\ref{se1}),
in the sense that it corresponds to a well-defined non-perturbative 
string theory. It is not even known how to make this statement precise.
On the other hand, in the case of equations (\ref{pe1}) and (\ref{Zexp1}),
the problem can be formulated easily, because a non-perturbatively 
defined zero-dimensional field theory is simply characterized by a 
potential which is bounded from below. This means that `correct' values of 
$\theta$ should be such that there exists $V_{\kappa, \theta}$ for which
\begin{equation}
\label{Znp1}
Z_{\theta}(\kappa)
= \int\! {dx\over\sqrt{2\pi}}\,e^{-x^{2}/2 -V_{\kappa,\theta}(x)} .
\end{equation}
We have not tried to solve directly (\ref{Znp1}) for $\theta$ and 
$V_{\kappa,\theta}$, but we will see that the non-linear $\sigma$ 
model approach yields unambiguously the potential
\begin{equation}
\label{solV}
V_{\kappa}(x) =\ln 2 +\sqrt{\kappa /8}\, x^{3} + \kappa x^{4}/16
\end{equation}
which corresponds to $\theta = 0$.
The $\ln 2$ term takes care of the two equivalent minima of
$V_{\kappa}$ that occur at $x=0$ and $x=-2\sqrt{2/\kappa}$. The fact 
that (\ref{solV}) is a solution relies on the nice identity
\begin{equation}
\label{id1}
\langle e^{-\sqrt{\kappa /8}\, x^{3} -\kappa 
x^{4}/16}\rangle_{\rm pert} = \langle e^{\kappa 
x^{4}/4}\rangle _{\rm pert}\, ,
\end{equation}
where $\langle\cdots\rangle_{\rm pert}$ is the 
perturbative expansion in $\kappa$ with $\langle x^{2}\rangle_{\rm 
pert}=1$. Remarkably, we will see in Section 3 that (\ref{id1}) can 
be generalized to the case of path integrals.

We have not proven that $\theta=0$ with the potential (\ref{solV}) is 
the only solution to (\ref{Znp1}), but we believe that this is likely 
to be the case, because identities like (\ref{id1}) are very 
peculiar. This illustrates the point \cite{fer1,fer2} that the 
gauge theory or non-linear $\sigma$ model integrals we start from 
yield the correct and probably unique non-perturbative definitions of 
the double scaled theories. This is an important point of principle, 
because even though the general arguments \cite{tHooft,DK} show that double
scaled matrix integrals reproduce the {\it perturbative expansion} of 
string theories, there might be a priori a distinction between 
non-perturbative string theory and non-perturbative matrix integral.

\section{Simple integrals}

We will consider the most general ${\rm O}(N)$ symmetric non-linear 
$\sigma$ model with a compact target space $\cal M$ of dimension $N$ and 
quadratic mass terms. By using 
cartesian coordinates $(x_{1},\ldots ,x_{N}, x_{N+1}=z)$, the
equation for the target space can be written in terms of a single 
function $f(z)$ as
\begin{equation}
\label{tseq}
{\cal M}:\ \sum_{i=1}^{N} x_{i}^{2} = {\vec x}^{2} = f(z)\, .
\end{equation}
We will limit our investigations to the cases where $f(z)$ is a polynomial.
At the North pole $z=z_{N}>0$ we have $f(z_{N}) = 0$ and $f'(z_{N})<0$. We 
will see below that for theories with critical points, there exists a 
$z<z_{N}$ for which $f'$ vanishes.
We will choose the coordinates $\vec x$ and $z$ such that
\begin{equation}
\label{condf}
f(0) = 1\, ,\quad f'(0)=0\, ,\quad f'(z)<0 \ {\rm for}\ z\in ]0,z_{N}]\, .
\end{equation}
The metric can be written in terms of the local inverse 
$F=f^{-1}$ of $f$ as
\begin{equation}
\label{metric}
g = \left(\delta_{ij} + 4 F'({\vec x}^{2})^{2} x_{i}x_{j}\right)\,
dx_{i}\otimes dx_{j}\, ,
\end{equation}
and the partition function is
\begin{equation}
\label{Zdef}
{\cal Z} = \int_{\cal M}\! {d^{N}\vec x\, \sqrt{\det g}\over
{\rm Vol}({\cal M})}\, e^{-N r {\vec x}^{2}/2}\, ,
\end{equation}
where $r$ is the mass parameter and ${\rm Vol}({\cal M})$ the volume 
of $\cal M$. The large $N$ limit will be taken at fixed $r$. By rescaling
the coordinates, we see that taking the large $r$ limit is 
equivalent to taking the large target space, or weak coupling, limit. The 
coupling constant for the theory is thus $1/r$. Equivalently, we can 
say that our models are asymptotically free, and the large mass limit 
corresponds to a weak coupling limit.
\subsection{Sphere}
Let us start with the simplest case corresponding to ${\cal M} = 
S^{N}$ and
\begin{equation}
\label{fsph}
f(z) = 1 - z^{2}\, .
\end{equation}
The first question one may ask is whether a proper `polymer' interpretation
of the theory can be given. This is not obvious, because the 
contribution of the metric in (\ref{Zdef}), which in the case of the 
sphere is \smash{$\sqrt{\det g} = 1/\sqrt{1-{\vec x}^{2}}$}, does not scale 
properly. However, this is due to a bad choice of coordinates, and it 
is possible to cast (\ref{Zdef}) in the form (\ref{vecint}).
To do that, we 
first introduce the spherical angle $\theta$, in terms of which
\begin{equation}
\label{Zst}
{\cal Z} = {\Gamma((N+1)/2)\over\sqrt{\pi}\, \Gamma (N/2)}
\int_{0}^{\pi}\! d\theta\, (\sin\theta)^{N-1} e^{-{Nr\over 2}
\sin^{2}\theta}\, .
\end{equation}
We then interpret $\theta$ as a radial coordinate emanating from the 
North (or South) pole for an $N$-dimensional plane $(V_{1},\ldots,V_{N})$,
$\theta = |\vec V|/\sqrt{r}=\rho /\sqrt{r}$. Physically, this
plane approximate the sphere 
for large $r$, and we can expect that the effects of the deviation 
from the sphere can be taken into account in an effective potential. 
Indeed, the partition function (\ref{Zst}) can be rewritten as
\begin{equation}
\label{Zpertpol}
{\cal Z} = {\Gamma ((N+1)/2) \over 2\pi^{(N+1)/2} r^{N/2}}
\int_{|\vec V|\leq \pi\sqrt{r}}\! d^{N}\vec V\, e^{-N U_{N}({\vec V}^{2})}
\end{equation}
with
\begin{equation}
\label{UNdef}
U_{N}(\rho^{2}) = {r\over 2}\sin^{2}(\rho /\sqrt{r}) - 
{N-1\over N} \ln {\sqrt{r}\sin (\rho /\sqrt{r})\over\rho} \,\cdotp
\end{equation}
One can then show straightforwardly that for the purposes of the double 
scaling limits, which involve the $N\rightarrow\infty$ limit, we can 
use the $N$-independent potential
\begin{equation}
\label{Uinf}
U(\rho^{2}) =  {r\over 2}\sin^{2}(\rho /\sqrt{r})-\ln {\sqrt{r}
\sin (\rho /\sqrt{r})\over\rho}
\end{equation}
and the integral
\begin{equation}
\label{intUinf}
\int_{|\vec V|\leq\pi\sqrt{r}}\! d^{N}\vec V\, e^{-NU({\vec V}^{2})}\, .
\end{equation}
We thus get a standard Feynman diagram interpretation, because the 
constraint $|\vec V|\leq\pi\sqrt{r}$ cannot be seen in perturbation theory.
We could then keep using (\ref{intUinf}) and study the critical point and 
double scaling limit. However, we prefer to present alternative 
derivations using (\ref{Zdef}) or (\ref{Zst}).

The integral (\ref{Zst}) can be explicitly calculated, by 
expanding $\exp (-(Nr/2)\sin^{2}\theta)$ in power series and performing the 
integrals using Euler $B$ function. This yields a series for a confluent 
hypergeometric function, and we get
\begin{equation}
\label{Zexact}
{\cal Z} = {}_{1}F_{1}(N/2,(N+1)/2;-Nr/2)=
e^{-Nr/2} {}_{1}F_{1} (1/2 , (N+1)/2 ; Nr/2)\, .
\end{equation}
The first expression is useful to obtain the asymptotic expansion 
at large $r$,
\begin{equation}
\label{Zpertexp}
{\cal Z} = {\Gamma ((N+1)/2)\over \sqrt{\pi} (Nr)^{N/2}}
\left[ \sum_{k=1}^{K} {\Gamma (N/2+k) \Gamma (k+1/2)\over\sqrt{\pi}\,
\Gamma (N/2) k!}\Bigl({2\over rN}\Bigr)^{k} + {\cal O}(1/r^{K+1})\right]\, ,
\end{equation}
while the second expression yields a convergent strong coupling 
expansion
\begin{equation}
\label{Zscexp}
{\cal Z} = {\Gamma ((N+1)/2)\over\sqrt{\pi}} \sum_{k=0}^{\infty} 
{\Gamma (k+1/2)\over k!\, \Gamma (k+(N+1)/2)} \Bigl({rN\over 2}\Bigr)^{k}\, .
\end{equation}
The large $N$ expansion can be obtained as usual from the 
perturbative expansion (\ref{Zpertexp}) by resumming the contributions 
at each order in $1/N$. It is convenient to introduce a rescaled partition 
function
\begin{equation}
\label{Zrdef}
\tilde {\cal Z} = \sqrt{1-1/r}\, {\sqrt{\pi} (Nr/2)^{N/2}\over
\Gamma ((N+1)/2)} \, {\cal Z}\, ,
\end{equation}
for which we have when $r>1$
\begin{equation}
\label{ZNexp}
\tilde {\cal Z} = 1 + {3\over 4 (r-1)^{2}N} +  \Bigl({5\over 2(r-1)^{3}} + 
{105\over 32 (r-1)^{4}}\Bigr){1\over N^{2}} + {\cal O}(1/N^{3})\, .
\end{equation}
We see that the $1/N$ expansion breaks down when $r\rightarrow 1^{+}$.
At $r=1$ we have a Kazakov critical point, and (\ref{ZNexp}) suggests
to consider the double scaling limit
\begin{equation}
\label{sca1}
N\rightarrow\infty\, ,\quad r\rightarrow 1^{+}\, ,\quad N(r-1)^{2} = 
1/\kappa = {\rm constant}.
\end{equation}
In such a limit, only the most singular, universal, terms in (\ref{ZNexp}) 
survive, and we get
\begin{equation}
\label{Zsca}
\tilde {\cal Z}\rightarrow Z = 1 + {3\over 4}\, \kappa +
{105\over 32}\,\kappa^{2} + {\cal O}(\kappa^{3})\, .
\end{equation}
To show that the scaling (\ref{sca1}) is consistent to all orders,
one can use directly (\ref{Zexact}) and check that the 
corresponding hypergeometric differential equation reduces to the 
`string equation' (\ref{pe1}). However, this 
method does not yield the value of $\theta$ in (\ref{zt1}). The most 
fruitful approach, that can be generalized to the case of path 
integrals (see for example \cite{fer3,fer4}), is to implement the 
constraint (\ref{tseq}) with a Lagrange multiplier $\alpha$ and then
perform explicitly the integral over $\vec x$. We obtain
\begin{equation}
\label{Zveff}
{\cal Z}\propto \int_{-\infty}^{+\infty}\! dz d\alpha\,
e^{-N v_{\rm eff}(z,\alpha)}\, ,
\end{equation}
with the effective potential
\begin{equation}
\label{veff}
v_{\rm eff}(z,\alpha) = {1 \over 2} (r-\alpha) f(z) + {1\over 2} 
\ln\alpha\, .
\end{equation}
We are not keeping track of the trivial prefactors, because they can be 
restored easily on the double scaled partition functions by using the
normalization condition $Z=1+{\cal O}(\kappa)$.
At large $N$, $\cal Z$ is dominated by the minima of $v_{\rm eff}$. 
It is straightforward to check that for $r>1$ (weak coupling) the 
stable saddle points are
\begin{equation}
\label{wcsp}
z_{*} = \pm\sqrt{1-1/r}\, , \quad \alpha_{*} = r\, , 
\end{equation}
while for $r<1$ (strong coupling) we have
\begin{equation}
\label{scsp}
z_{*} = 0\, ,\quad \alpha_{*} = 1\, .
\end{equation}
We recover the critical point at $r=1$, where a 
transition corresponding to the merging of the two weakly coupled 
saddle points occurs.\footnote{Note that the ${\mathbb Z}_{2}$
symmetry $z\mapsto -z$ is never broken, because in zero (or one) 
dimension we have to sum over all the saddle points. Symmetry 
breaking does occur at weak coupling for the two-dimensional version of 
the model, see Section 3 and \cite{fer3,fer4}, and the Kazakov critical 
point corresponds to a genuine phase transition in that case.} Since 
we are interested in the vicinity of the critical point only, and the 
critical variable is $z$, we can integrate over $\alpha$ by using 
the equation $\partial v_{\rm eff}/\partial\alpha = 0$ which yields 
$\alpha = 1/f(z)$. By rescaling $z\rightarrow z/N^{1/4}$ and 
expanding the potential (\ref{veff}) around $z=0$, we get
\begin{equation}
\label{Zv2}
{\cal Z}\propto \int_{-\infty}^{+\infty}\! dz\,
e^{\sqrt{N} (r-1) z^{2}/2 - z^{4}/4 + {\cal O}(1/\sqrt{N})}\, .
\end{equation}
We then immediately see that the scaling (\ref{sca1}) yields
\begin{equation}
\label{Zex1}
Z(\kappa)= {e^{-1/(4\kappa)}\over \sqrt{4\pi}}\int_{-\infty}^{+\infty}\!
dx\, e^{x^{2}/2 - \kappa x^{4}/4}\, ,
\end{equation}
where we have used the variable $x=\kappa^{1/4} z$ and we have 
restored the prefactors. The formula (\ref{Znp1}) with the potential
(\ref{solV}) is obtained by substituting 
$x\rightarrow 1/\sqrt{\kappa} + \sqrt{2}\, x$.

As opposed to the case of ordinary vector or matrix models, our partition 
function is perfectly well-defined at strong coupling
and we can go through the critical point at $r=1$. Equation (\ref{Zv2}) is 
actually valid both for $r>1$ and for $r<1$, and it shows that we can 
consider a double scaling limit from strong coupling,
\begin{equation}
\label{sca1sc}
N\rightarrow\infty\, ,\quad r\rightarrow 1^{-}\, , \quad
N(r-1)^{2} = 1/\kappa = {\rm constant}\, ,
\end{equation}
yielding a ``dual'' double scaled partition function
\begin{equation}
\label{Zd}
Z_{\rm D}(\kappa) = \int_{-\infty}^{+\infty}\! {dx\over\sqrt{2\pi}}
e^{-x^{2}/2 - \kappa x^{4}/4}\, .
\end{equation}
Of course, $\cal Z$  does not have a Feynman diagram expansion for 
$r<1$ and thus $Z_{\rm D}$ does not have an interpretation in terms
of `polymers.' However, the very existence of $Z_{\rm D}$, which relies on 
the fact that our non-linear $\sigma$ model is non-perturbatively defined, 
has some interesting consequences.
The common origin (\ref{Zv2}) of the weak-coupling (\ref{Zex1}) and strong 
coupling (\ref{Zd}) partition functions implies that $\int\exp (x^{2}/2 - 
\kappa x^{4}/4) dx$ and $\int\exp (-x^{2}/2 - \kappa x^{4}/4) dx$ satisfy 
the same `string' equation
\begin{equation}
\label{Zddiff}
16\kappa^{2}\, y'' + 4(8\kappa +1)\, y' + 3\, y = 
0\, .
\end{equation}
The Borel summable, alternate series solution of (\ref{Zddiff}) yields 
$Z_{\rm D}$ while the non-Borel summable solution yields 
$\exp(1/(4\kappa))\, Z$. $Z$ itself actually satisfies (\ref{Zddiff}) with 
$\kappa\rightarrow -\kappa$, see (\ref{pe1}), and thus the equation 
(\ref{Zd}) immediately implies the identity (\ref{id1}).
\vfill\break
\subsection{Multicritical metrics}
Let us now study the case of a general metric (\ref{metric}).
Equation (\ref{Zveff}) is replaced by
\begin{equation}
\label{Zveffgen}
{\cal Z}\propto \int_{-\infty}^{+\infty}\! dz d\alpha\,
e^{-N v_{\rm eff}(z,\alpha) + {1\over 2} \ln (4f(z) + f'(z)^{2})}\, ,
\end{equation}
with the potential (\ref{veff}). The saddle point equations read
\begin{equation}
\label{spgen}
\alpha = 1/f(z)\, ,\quad (r-\alpha) f'(z) = 0\, .
\end{equation}
The analysis is then very similar to the case of the sphere, equations 
(\ref{wcsp}) and (\ref{scsp}).
At weak coupling, the stable saddle point is\footnote{There can be 
several saddle points, given by the different branches of $f^{-1}$, as in 
(\ref{wcsp}), but this is irrelevant for our purposes.}
\begin{equation}
\label{wcspgen}
z_{*} = f^{-1}(1/r) = F(1/r)\, ,\quad \alpha_{*} = r\, .
\end{equation}
This is valid as long as $r>1$. Because on the conditions on $f$ 
listed in
(\ref{condf}), a critical point occurs at $r=1$ and the stable saddle
point for $r<1$ is given by (\ref{scsp}). The critical variable being $z$, 
we can integrate out $\alpha$ for the purposes of the double 
scaling limits. The term $\ln (4f + f'^{2})$ is also irrelevant in 
these limits. We can thus work with
\begin{equation}
\label{Zsim}
\int_{-\infty}^{+\infty}\! dz\, e^{-N v(z;r)}\, ,
\end{equation}
where the potential is
\begin{equation}
\label{vgendef}
v(z;r) = {r\over 2} f(z) - {1\over 2} \ln f(z)\, .
\end{equation}
We have $v(z;r=1) = v(0,r=1) + {\cal O}(z^{p})$. 
It turns out that only even values of $p\geq 4$ can be obtained.
The $m^{\rm th}$-critical point, $m\geq 2$, is
defined by the condition $v(z;r=1) = v(0;r=1) + {\cal O}(z^{2m})$. 
It corresponds to the choice
\begin{equation}
\label{critmet}
f_{m}(z) = 1 - a_{m} z^{m}\, ,
\end{equation}
where $a_{m}$ is an arbitrary positive real number. Relevant deformations 
are defined to be the 
perturbations of $f_{m}$ that generate terms of order $z^{k}$ for $k\leq 
2m -1$ in the potential $v$. A priori one may want to consider
\begin{equation}
\label{fpert}
f(z) = f_{m}(z) - z^{2}\sum_{k=0}^{2m-3} \epsilon_{k} z^{k}\, ,
\end{equation}
but it is easily checked that only the $\epsilon_{k}$ for $0\leq 
k\leq m-3$ can survive in a consistent double scaling limit.
Together with $\delta = r-1$, we thus have $m-1$ relevant operators at 
the $m^{\rm th}$ order critical point. The correct scaling is
\begin{equation}
\label{scagen}
N\rightarrow\infty\, ,\quad N\delta ^{2}= {\rm cst}= 1/\kappa\, ,\quad
N^{{m-k-2\over 2m}}\epsilon_{k} = {\rm cst}\propto t_{k}\, .
\end{equation}
With a suitable normalization for the $t_{k}$s, and by defining 
$t_{m-2}=1$, the $m^{\rm th}$ order double scaled partition function can 
then be written as
\begin{equation}
\label{Zm}
Z_{m} \propto \int_{-\infty}^{+\infty}\! dx\, \exp\Bigl( \sum_{k=0}^{m-2} 
t_{k}\, x^{k+2} - \kappa\sum_{k=0}^{2m-4}\!\!\!
\sum_{\scriptstyle k_{1}+k_{2}= k\atop\scriptstyle 0\leq 
k_{1},k_{2}\leq m-2}\!\!\!\! t_{k_{1}}t_{k_{2}}\, x^{k+4} \Bigr)\, .
\end{equation}
The `coupling constant' $\kappa$ could of course be absorbed in the 
definition of the $t_{k}$s. It is singled out by the fact that it is the 
most relevant deformation, as the scaling (\ref{scagen}) shows.
For $m=2$, we recover (\ref{Zex1}), while for higher criticality we 
obtain, for example,
\begin{equation}
\label{ex1}
Z_{3}\propto \int_{-\infty}^{+\infty}\! dx\, e^{t_{0}x^{2} + x^{3} - 
\kappa (t_{0}^{2}x^{4} + 2 t_{0} x^{5} + x^{6})}\, ,
\end{equation}
\begin{equation}
\label{ex2}
Z_{4}\propto \int_{-\infty}^{+\infty}\! dx\, e^{t_{0}x^{2} + t_{1}x^{3} + 
x^{4} - \kappa (t_{0}^{2}x^{4} + 2t_{0}t_{1}x^{5} + (t_{1}^{2} + 
2t_{0})x^{6} + 2t_{1} x^{7} + x^{8})}\, , \quad {\rm etc\ldots}
\end{equation}
From (\ref{Zm}), all correlators can of course be calculated, by taking 
derivatives with respect to the parameters. One can also study positivity 
by looking at the partition functions when only $\kappa$ is turned on. The 
formula generalizing (\ref{Zex1}) is
\begin{equation}
\label{Zex2}
Z_{m}(\kappa) = {m e^{-1/(4\kappa)}\over 2^{(1+(-1)^{m})/2}\sqrt{2\pi} 
(2\kappa)^{1/2 - 1/m}} \int_{-\infty}^{+\infty}\! dx\, e^{x^{m} - \kappa 
x^{2m}}\, ,
\end{equation}
where the normalization is chosen such that $Z_{m} = 1 + {\cal 
O}(\kappa)$. The `string equation' is
\begin{equation}
\label{sem}
4 m \kappa^{2}\, Z''_{m} + \left(\strut 2(5m-2)\kappa -m\right)\, 
Z'_{m} + {(2m-1)(m-1)\over m}\, Z_{m} = 0\, ,
\end{equation}
from which one can deduce the perturbative expansion of $Z_{m}$, for which 
all the coefficients are positive. The same is true for $\ln Z_{m}$, 
as required by a correct statistical polymer
interpretation. This is manifest from equation (\ref{id1}) in the case 
$m=2$, and can be proven for general $m$ from the differential 
equation. Finally, let us note that it is 
possible to evaluate explicitly the integrals (\ref{Zex2}). The idea is to 
perform a strong coupling expansion at large $\kappa$. From experience, we 
know that such expansions are often convergent. The expansion is obtained 
by rescaling $x^{2m}\rightarrow x^{2m}/\kappa$ in (\ref{Zex2}) and 
expanding the exponential in powers of $1/\sqrt{\kappa}$. The resulting 
integrals are elementary, and the final result is a hypergeometric series. 
We end up with
\begin{equation}
\label{exfor1}
Z_{m}(\kappa) = {\Gamma(1/(2m))\, e^{-1/(4\kappa)}\over
2^{1-1/m +(-1)^{m}/2}\sqrt{2\pi} \kappa^{1/2 - 1/(2m)}}\,
{}_{1}F_{1} (1/(2m),1/2;1/(4\kappa))
\end{equation}
for $m$ odd and
\begin{eqnarray}
&& \hskip -1.5cm Z_{m}(\kappa) = {e^{-1/(4\kappa)}\over
2^{1-1/m +(-1)^{m}/2}\sqrt{2\pi} \kappa^{1/2 - 1/(2m)}}\,\Bigl(
\Gamma(1/(2m)) \, {}_{1}F_{1} (1/(2m),1/2;1/(4\kappa))\nonumber\\
&&\hskip 3.5cm + {\Gamma (1/2 + 1/(2m))\over\sqrt{\kappa}} \,
{}_{1}F_{1}(1/2 + 1/(2m), 3/2; 1/(4\kappa)) \Bigr)
\label{exfor2}\\ \nonumber
\end{eqnarray}
for $m$ even. In the special case $m=2$, there is a relation between the 
confluent hypergeometric functions appearing in (\ref{exfor2}) and the 
Bessel functions, and we recover (\ref{zt1}) with $\theta = 0$.

\section{Path integrals}

The results of the previous Section can be generalized to the case of 
path integrals. For the sake of brevity, we will consider only the lower 
critical point corresponding to a sphere target space. In one dimension, 
the resulting model is the quantum version of a famous integrable 
mechanical problem first studied by C.~Neumann \cite{neumann}: the motion 
of a particle of mass $m$ on the $N$ dimensional sphere of radius $a$, with 
a quadratic potential characterized by the pulsation $\omega$. Quantum 
mechanically, there are two regimes. When the sphere is very large, the 
effects of the curvature are negligible, and the problem is well 
approximated by the $N$ dimensional harmonic oscillator. This weakly 
coupled regime is valid as long as the harmonic oscillator wave functions 
extend on a distance much smaller that $a$, that is
\begin{equation}
\label{wccond}
r = {m\omega a^{2}\over N\hbar } \gg 1\, .
\end{equation}
On the other hand, when $r\ll 1$, the hamiltonian reduces to the exactly 
solvable rigid rotator, around which a strong coupling expansion can be 
performed. At the transition between these two qualitatively different 
regimes lies the critical point we will use to define the double scaling 
limit. In suitable units, the euclidean
partition function that generalizes (\ref{Zdef}) is
\begin{equation}
\label{Zdef2}
{\cal Z}_{T} = \int\! \left[ d^{N}\vec x(t)\, \sqrt{\det g (\vec x(t))}\right]
\, \exp \left[ - {Nr\over 2} \int_{-T/2}^{T/2}\! dt \Bigl( g_{ij}\,
{dx^{i}\over dt}{dx^{j}\over dt} + {\vec x}^{2} \Bigr)\right]\, ,
\end{equation}
where the path integral measure is normalized such that the ground state 
energy $E$ for the hamiltonian
\begin{equation}
\label{Hdef}
H = {{\vec L}^{2}\over 2N^{2}} + {r^{2}\over 2} {\vec x}^{2}\, ,
\end{equation}
where $-{\vec L}^{2}$ is the Laplacian on the $N$-sphere, is given by
\begin{equation}
\label{ener}
E = -{r\over N}\lim_{T\rightarrow\infty}{\ln {\cal Z}_{T}\over T}\,\cdotp
\end{equation}
It is interesting to note that, as in the case of the zero-dimensional 
integrals (equations (\ref{Zpertpol}) and (\ref{UNdef})), the
perturbation theory in $1/r$ for the quantum mechanical 
non-linear $\sigma$ model (\ref{Zdef2}) is reproduced by a linear $\sigma$ 
model with a suitable potential that encodes the effects of the curvature 
of the sphere. By using the coordinates $\vec V$ defined at the beginning 
of Section 2.1, and by rescaling the wave functions 
$\psi\rightarrow\psi / (\sin\theta)^{(N-1)/2}$, the Schr\"odinger equation 
for the ground state of (\ref{Hdef}) can indeed be cast in the form
\begin{equation}
\label{Scheq}
-{1\over 2N^{2}}\, \Delta\psi + U_{N}(|\vec V|)\,\psi = {E\over 
r}\,\psi\, , 
\end{equation}
where now $\Delta$ is the flat $N$-dimensional Laplacian and
\begin{equation}
\label{UNdef2}
U_{N}(\rho) = -{N-1\over 4rN^{2}} +
{(N-3)(N-1)\over 8N^{2}} \Bigl( {1\over r\tan^{2} (\rho 
/\sqrt{r})} - {1\over\rho} \Bigr) + {r\over 2} \sin^{2} (\rho /\sqrt{r})\, .
\end{equation}
As a side remark, let us note that the formulation given by the equations 
(\ref{Scheq}) and (\ref{UNdef2}), in addition to providing a consistent 
polymer interpretation in the double scaling limit,
allows one to evaluate, using standard linear $\sigma$ model techniques, 
the large order behaviour of perturbation theory for a non-linear 
$\sigma$ model. As we will see later, the perturbation series is not Borel 
summable. From the linear $\sigma$ model point of view, the
non-perturbative contributions depend on the boundary conditions at 
$\rho = \pi\sqrt{r}$ that must be imposed to recover the full non-linear 
model. This is particularly obvious in the case $N=1$, which can be solved 
exactly in terms of Mathieu functions. In general, by taking into account
the exponential decrease of the harmonic
oscillator wave functions, we see that the non-perturbative contributions 
are of order $\exp (-N\pi^{2} r/2)$, which is an instanton effect from the 
non-linear $\sigma$ model point of view.

Equation (\ref{Scheq}) shows that the large $N$ limit is a semi-classical 
limit, since $N^{2}$ plays the r\^ole of $\hbar$. We could use this 
idea to study the $1/N$ expansion. However, it is more convenient to use 
the Lagrange multiplier method, as in the case of the zero dimensional 
integrals (\ref{Zveff}). We obtain
\begin{equation}
\label{lag2}
{\cal Z}\propto \int\! \left[ dz(t)d\alpha (t)\right]\, e^{-N s_{\rm 
eff}[z,\alpha]}\, ,
\end{equation}
with the effective action
\begin{equation}
\label{seff}
s_{\rm eff}[z,\alpha] = {r\over 2} \int\! dt\, \Bigl[ \Bigl({dz\over 
dt}\Bigr)^{2} + (\alpha - 1)z^{2} - \alpha \Bigr] + {1\over 2} \tr\ln 
\Bigl( - {d^{2}\over dt^{2}} + \alpha \Bigr)\, .
\end{equation}
The saddle points can be deduced from the effective potential which is 
derived from (\ref{seff}) by taking $\alpha$ constant,
\begin{equation}
\label{veff1}
v_{\rm eff} = {r\over 2} \Bigl[ (\alpha -1)z^{2} - \alpha\Bigr] + 
{\sqrt{\alpha}\over 2}\, \cdotp
\end{equation}
For $r>1/2$ (weak coupling) we get
\begin{equation}
\label{spwc1}
z_{*} = \pm\sqrt{1-1/(2r)}\, ,\quad \alpha_{*} = 1\, ,
\end{equation}
while for $r<1/2$ (strong coupling) we have
\begin{equation}
\label{spsc1}
z_{*} = 0\, ,\quad \alpha_{*} = 1/(4r^{2})\, .
\end{equation}
The critical point occurs at $r=1/2$. To perform explicitly the double 
scaling limit, we introduce the rescaled variables
\begin{equation}
\label{scaf}
\tau = N^{-1/3} t\, ,\quad x = N^{1/3} z\, ,\quad \beta = N^{2/3} 
(\alpha -1)\, ,\quad \kappa^{-1} = N (r-1/2)^{3/2}\, ,
\end{equation}
in terms of which
\begin{equation}
\label{Zd11}
Ns_{\rm eff} = {1\over 4} \int\! d\tau \, \Biggl[ \Bigl( {dx\over 
d\tau}\Bigr)^{2} + \beta (x^{2} - 2\kappa^{-2/3}) - {\beta^{2}\over 
4}\Biggr] + {\cal O}(1/N^{2/3})\, .
\end{equation}
Rescaling $\tau\rightarrow\kappa^{1/3}\tau /\sqrt{2}$ and $x\rightarrow 
2^{1/4}\kappa^{1/6} x$, we see that the double scaling limit
\begin{equation}
\label{dsca2}
N\rightarrow\infty\, ,\quad r\rightarrow 1/2^{+}\, ,\quad 
N(r-1/2)^{3/2} = 1/\kappa = {\rm constant}
\end{equation}
yields the double scaled partition function
\begin{equation}
\label{Zsca2}
Z(\kappa) = \int\! [dx(t)]\, \exp\left[ -\int\! d\tau\, \Biggl(
{1\over 2} \Bigl({dx\over d\tau}\Bigr)^{2} - x^{2} + {\kappa 
x^{4}\over 2\sqrt{2}} + 
{1\over\sqrt{2}\kappa} \Biggr) \right]\, ,
\end{equation}
where we have integrated out the auxiliary field $\beta$ and
the $\kappa$-independent measure $[dx]$ is normalized such 
that $Z = 1 +{\cal O}(\kappa)$ as usual. Equation (\ref{Zsca2}) is the 
one-dimensional version of equation (\ref{Zex1}).

There remains to check our main point, that the perturbative 
expansion of (\ref{Zsca2}), or more precisely of $W = \ln Z$, has 
only positive coefficients and is not Borel summable. The perturbative 
series is obtained by expanding around the two equivalent minima of 
the potential $V(x) = -x^{2}+\kappa x^{4}/(2\sqrt{2})$. After a shift of
the variable $x$, we obtain the one-dimensional version of 
(\ref{Znp1}) and (\ref{solV}),
\begin{equation}
\label{Zsca2p}
Z(\kappa) = \int\! [dx(t)]\, \exp\left[ -\int\! d\tau\, \Biggl(
{1\over 2} \Bigl({dx\over d\tau}\Bigr)^{2} + 2x^{2}\Bigl( 
1-{\sqrt{\kappa}\over 2^{5/4}} x\Bigr)^{2} \Biggr) \right]\, .
\end{equation}
In zero dimensions, positivity was obvious thank's to the identity 
(\ref{id1}). Remarkably, a similar and highly non-trivial
identity exists in one dimension as well \cite{id1D}. It reads
\begin{equation}
\label{id2}
Z(\kappa)\, \mathop{=}_{\rm pert}\,
\Bigl\langle\exp\Bigl( {\kappa\over 2^{7/2}}\int\! 
d\tau\, |z|^{4}\Bigr)\Bigr\rangle_{\rm pert} \, ,
\end{equation}
where $z=z_{1}+iz_{2}$ is a complex coordinate, and the 
average $\langle\cdots\rangle_{\rm pert}$ is defined by the 
canonically normalized propagator $\langle z_{i}(\tau) 
z_{j}(0)\rangle_{\rm pert} =  e^{-|\tau |}\, \delta_{ij}/2$.
The equality is valid to 
all orders of perturbation theory. The left hand side can be viewed as 
giving the probably unique field theoretic non-perturbative definition of 
the manifestly positive and non Borel summable partition function on
the right hand side.

One could go further and study multicritical metrics as in Section 2. We 
will let this exercise to the reader, and rather close this paper with a 
two dimensional example. The partition function we start from, that 
replaces (\ref{Zdef}) and (\ref{Zdef2}), is
\begin{equation}
\label{Zdef3}
{\cal Z} = \int\! \left[ d^{N}\vec x(\sigma)\, 
\sqrt{\det g (\vec x(\sigma))}\right]
\, \exp \left[ - {N\over 2g^{2}} \int\! d^{2}\sigma \Bigl( g_{ij}\,
\partial_{a} x^{i} \partial_{a} x^{j} +
m^{2}{\vec x}^{2} \Bigr)\right]\, .
\end{equation}
The theory needs to be renormalized, and quantum mechanically the 
dimensionless coupling $g$ is replaced by a mass scale $\Lambda$. The only 
dimensionless parameter is then $r=m^{2}/\Lambda^{2}$. One can give 
straightforwardly a `polymer' interpretation to (\ref{Zdef3}), by using 
dimensional regularization, because in that case the non-trivial factor in 
the path integral measure drops out, and the other terms have automatically 
the correct 't Hooft scaling. This is unlike the $D=0$ or $D=1$ cases 
discussed previously, for which a reformulation in terms of a linear 
$\sigma$ model was needed. One can then show that there is a critical
point at $r=1$, and that in the double scaling limit
\begin{equation}
\label{sca2D}
N\rightarrow\infty\, ,\ \delta = r-1\rightarrow 0^{+}\, ,
\ N\delta - 3\ln N = {\rm constant} = {1\over 2\kappa} + 
3\ln\kappa\, ,\ \sigma^{a} = \sqrt{N}\, x^{a}\, ,
\end{equation}
the partition function reduces to
\begin{equation}
\label{Z2D}
Z(\kappa) = \int\! [d\phi(x)]\, \exp\Biggl[ -\int\! d^{2}x\, \Bigl(
{1\over 2}\, \partial_{a}\phi\partial_{a}\phi - {\Lambda^{2}\over 
4\kappa} :\!\phi^{2}\! : +\pi\Lambda^{2} :\!\phi^{4}\! :\Bigr)
\Biggr]\, .
\end{equation}
The proof can be found in \cite{fer3}. The $\ln N$ correction to the 
na\"\i ve scaling in (\ref{sca2D}) is reminiscent of the $c=1$ matrix 
model \cite{c1}. The normal ordering is defined at the scale $\mu = 
\Lambda/\sqrt{\kappa}$ so that there is no tadpole in perturbation 
theory. The perturbative expansion is indeed obtained after shifting 
the field $\phi\rightarrow 1/\sqrt{8\pi\kappa} + \phi$ and writing 
(\ref{Z2D}) in the form
\begin{equation}
\label{Z2Db}
Z(\kappa) = \int\! [d\phi(x)]\, \exp\Biggl[ -\int\! d^{2}x\, \Bigl(
{1\over 2}\, \partial_{a}\phi\partial_{a}\phi + {\Lambda^{2}\over 
2\kappa} :\!\phi^{2}\! :+ \Lambda^{2}\sqrt{2\pi\over\kappa}\, 
:\!\phi^{3}\!: +\pi\Lambda^{2} :\!\phi^{4}\! :\Bigr)
\Biggr]\, .
\end{equation}
We have not been able to prove that the coefficients of the expansion 
of $\ln Z$ in powers of $\kappa$ are positive, because we don't know 
a two dimensional analogue of the identities (\ref{id1}) and (\ref{id2}). 
We conjecture that this is the case, and that the series for $\ln Z$ 
is not Borel summable. We have checked explicitly
the positivity of the first two coefficients (see Figure 1), 
with the result
\begin{equation}
\label{Z2D3l}
\lim_{V\rightarrow\infty} {\ln Z(\kappa)\over V\Lambda^{2}} = 0.2798 + 
0.2078\,\kappa + {\cal O}(\kappa^{2})\, ,
\end{equation}
where $V$ is the volume of the two dimensional space-time. 
The coefficient of order $k$ grows like $k!$ and probably
becomes more and more positive for $k\geq 2$.
The formula (\ref{Z2D}) nevertheless 
yields a non-perturbative definition of the non-Borel summable sum, 
as in all the examples that we have studied in the present paper.

\begin{figure}
\centerline{\epsfig{file=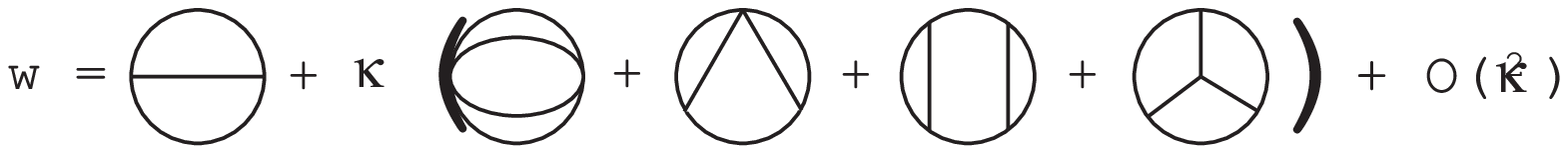,width=14cm}}
\caption{Feynman diagrams contributing to the double scaled partition 
function $\lim_{V\rightarrow\infty}
(\ln Z)/(V\Lambda^{2})$ up to terms of order $\kappa$.
\label{figu}}
\end{figure}
\section*{Acknowledgements}
I would like to acknowledge discussions with V.A.~Kazakov. This work 
was supported in part by a Robert H.~Dicke fellowship and by the Swiss 
National Science Foundation.
\vfill\eject

\begin{thebibliography}{99}
%
%\addcontentsline{toc}{section}{References}
%
\bibitem{polbook}{J.~Polchinski, {\it String Theory,} two volumes, 
Cambridge University Press 1998.}
%
\bibitem{BIPZ}{\'E.~Br\'ezin, C.~Itzykson, G.~Parisi and J.-B.~Zuber,
\cmp{59}{1978}{35}.}
%
\bibitem{BK}{\'E.~Br\'ezin and V.A.~Kazakov, \plb{236}{1990}{144},\\
M.R.~Douglas and S.~Shenker, \npb{355}{1990}{635},\\
D.J.~Gross and A.A.~Migdal, \prl{64}{1990}{127}.}
%
\bibitem{DK}{F.~David, \npb{257}{1985}{45},\\
V.A.~Kazakov, \plb{150}{1985}{282},\\
J.~Ambj\"orn, B.~Durhuus and J.~Fr\"ohlich, \npb{257}{1985}{433}.}
%
\bibitem{tHooft}{G.~'t~Hooft, \npb{72}{1974}{461}.}
%
\bibitem{rev1}{P.~Di Francesco, P.~Ginsparg and J.~Zinn-Justin,
\pr{254}{1995}{1}.}
%
\bibitem{rev2}{I.R.~Klebanov, {\it String theory in two dimensions,}
Trieste spring school in String Theory and Quantum Gravity 1991, 
hep-th/9108019.}
%
\bibitem{rev3}{\'E.~Br\'ezin and S.R.~Wadia editors, 
{\it The large $N$ expansion
in quantum field theory and statistical physics,} World Scientific 1993.}
%
\bibitem{fer1}{F.~Ferrari, \npb{617}{2001}{348}.}
%
\bibitem{fer2}{F.~Ferrari, {\it Large $N$ and double scaling limits in two 
dimensions,} NEIP-01-008, PUPT-1997, LPTENS-01/11, hep-th/0202002.}
%
\bibitem{polym}{J. Ambj\o rn, B. Durhuus, and T. J\' onsson, 
\plb{244}{1990}{403},\\
S. Nishigaki and T. Yoneya, \npb{348}{1991}{787},\\
P. Di Vecchia, M. Kato, and N. Ohta, \npb{357}{1991}{495},\\
G.W.~Semenoff and R.J.~Szabo, \mpla{11}{1996}{1185}.}
%
\bibitem{polymhD}{J.~Zinn-Justin, \plb{257}{1991}{335},\\
P.~Di Vecchia, M.~Kato, and N.~Ohta, \ijmpa{7}{1992}{1391},\\
G.~Eyal, M.~Moshe, S.~Nishigaki and J.~Zinn-Justin,
\npb{470}{1996}{369}.}
%
\bibitem{polymrev}{M.~Moshe, {\it Quantum field theory in singular limits,}
Les Houches lecture 1997, hep-th/9812029,\\
J.~Zinn-Justin, {\it Vector models in the large $N$ limit: a few 
applications,} lectures at the 11$^{\rm th}$ Taiwan Spring School on 
Particles and Fields 1997, hep-th/9810198.}
%
\bibitem{fermat}{G.W.~Semenoff and R.J.~Szabo, \ijmpa{12}{1997}{2135}.}
%
\bibitem{fer3}{F.~Ferrari, \plb{496}{2000}{212},\\
F.~Ferrari, \jhep{6}{2001}{57}.}
%
\bibitem{fer4}{F.~Ferrari, \npb{612}{2001}{151}.}
%
\bibitem{AD}{P.C.~Argyres and M.R.~Douglas, \npb{448}{1995}{93}.}
%
\bibitem{neumann}{C. Neumann, {\it Journal f\" ur die Reine und
Angewandte Mathematik} {\bf 56} (1859) 46.}
%
\bibitem{id1D}{R.~Seznec and J.~Zinn-Justin, \jmp{20}{1979}{1398}, \\
J.~Zinn-Justin, \jmp{22}{1981}{511},\\
R.~Damburg, R.~Propin and V.~Martyshchenko, \jpa{17}{1984}{3493}.}
%
\bibitem{c1}{\'E.~Br\'ezin, C.~Itzykson, G.~Parisi, J.-B.~Zuber, 
\cmp{59}{1978}{35},\\
\'E.~Br\'ezin, V.A.~Kazakov and Al.B.~Zamolodchikov, 
\npb{338}{1990}{673},\\
D.~Gross and M.~Milkovic, \plb{238}{1990}{217},\\
G.~Parisi, \plb{238}{1990}{209},\\
P.~Ginsparg and J.~Zinn-Justin, \plb{240}{1990}{333}.}
%

%
\end{thebibliography}
\end{document}